\documentstyle[sprocl]{article}

\def\be{\begin{equation}}
\def\ee{\end{equation}}
\def\bea{\begin{eqnarray}}
\def\eea{\end{eqnarray}}

\begin{document}

\begin{flushright}
\vspace{-3cm}

KUCP-0112\\
January 1998
\end{flushright}

\vspace{3cm}

\title{CONCLUDING REMARKS\break
---International Symposium on ``QCD Corrections and New Physics", 
October 27-29, 1997, Hiroshima, Japan---}
\author{Satoshi MATSUDA}
\address{Department of Fundamental Sciences, FIHS, Kyoto University 
\\ Yoshida, Sakyo-ku, KYOTO 606-01, JAPAN
\\E-mail: matsuda@phys.h.kyoto-u.ac.jp}
\maketitle

\vspace{3cm}

\abstracts{
This is a report of the concluding talk presented at the International 
Symposium on QCD Corrections and New Physics which was held during 
October 27-29, 1997 at Mielparque Hiroshima, Hiroshima, Japan. 
The report will be published in the Conference Proceedings.
}


\newpage
\title{CONCLUDING REMARKS}
\author{Satoshi MATSUDA}
\address{Department of Fundamental Sciences, FIHS, Kyoto University 
\\ Yoshida, Sakyo-ku, KYOTO 606-01, JAPAN
\\E-mail: matsuda@phys.h.kyoto-u.ac.jp}
\maketitle



I am not so sure that my talk as a last speaker 
will be that appropriate as ``Concluding Remarks" of the conference 
under the title \lq\lq QCD Corrections and New Physics\rq\rq. 
But I hope that from the point of view of \lq New Physics\rq\  
it will add some flavors to what we have discussed 
for the last three days.

Now, my role is supposed to conclude this three-day Hiroshima 
International Symposium, October 27-29, 1997, 
which was planned to get together 
people from abroad as well as from domestic institutes, 
who have been actively engaged in the study of 
``QCD Corrections and 
New Physics".

Here is the programme of the Symposium. 
The meeting has been quite successful, 
more than anticipated, mainly due to the active and cooprerative 
participation of all attendants to the intensive activities 
of the Symposium.
It has provided us participants  
opportunities of discussing physics face to face.

The topics covered in the conference are: Recent Results from 
FNAL, LEP, HERA, Aspects of QCD Theory, 
Structure Functions, Jets, Drell-Yan Processes, 
Top Quark Physics, Spin Physics, Weak Decays in QCD, 
Lattice Gauge Theory, and some others.
Future projects have been discussed as well.

I would say that in this meeting these topics have been studied 
and discussed in a serious and faithful manner 
with established theoretical tools and 
concrete experimental results. 
No hand waving arguments, but all serious discussions.

Let me summarize in a few words the future prospect of QCD. 
During the last two days and today we have heard some talks on 
\lq Spin Effects\rq: 
Chiral Odd Structure Functions in Drell-Yan, 
Polarization Effects in Top Quark Decays, 
Spin Physics in Top Pair Production, and so on. 
We have also had talks on \lq Resummation\rq\  of 
how \lq large $logs$\rq\ become relevant in various aspects of 
QCD theory on top of higher order effects 
of the QCD coupling constant. 
My personal view which I have had so far is that 
\lq{\bf Polarization}\rq\ 
and \lq{\bf Resummation of Large $logs$}\rq\ 
may be the key topics of QCD physics 
in the near future anyway, 
while \lq New Physics\rq\  might lie quite far away from us. 
Nevertheless we shall continue on and on to pursue 
something unexpectd, something beyond our present knowledge 
to achieve some great findings of \lq New Physcis\rq.

At this point please allow me to express my personal view on 
the appearance of some \lq New Physics\rq. 
In my mind I have had a feeling that 
\lq{\bf Supersymmetry}\rq\  might play a key role 
to the next big step of particle physics developments. 
I presume that most people here in the audience may have 
a similar anticipation.

Supersymmtery is a symmetry of very high quality. 
My belief is that nature favors high-bred characters of symmtery.
Nature seems to run after good looks of symmetry figures 
having a light touch of symmetry-breaking.
If we look back at the history of physics, we could easily count 
a few examples of how nature likes the beauties of 
simple symmetry.

In Classical Newtonian Mechanics we have a central force potential 
$V(r)=k r^n$ with $k$ and $n$ some constants. 
We have an infinite variety of choices for dynamical forces 
specified by the power $n$. 
However, nature seems to exhibit the specific taste of choosing 
the only two values of $n=-1$ and $n=2$ out of infinite choices, 
the values of very high symmetry 
which actually provide \lq closed orbits\rq\  of 
mass points under the central potentials. 
The first choice $n=-1$ dictates the motion of planets 
in classical theory, 
while in quantum theory it describes 
the dynamics of electrons inside atoms. 
The second one $n=2$ corresponding to a string force or an oscillator, 
explains photon quanta, oscillations of atoms in the crystals, 
or describes universally mass oscillations 
around stable points in nature ranging 
from the microscopic to the macroscopic world.

In Quantum Chromodynamics (QCD) the value $n=1$ seems 
to be chosen by the quark confining potential 
between quarks inside hadrons.
For example, consider mesons made up of a quark and an anti-quark 
with a distance separation $L$. 
Suppose the potential between them is given by 
$V(L)=\sigma L^n$. 
Then the total Hamiltonian of the system 
with quark mass $\mu$ will be for $p\gg\mu c$ 
\[
H=2\sqrt{p^2c^2+\mu^2c^4}+V(L)\approx 2pc+\sigma L^n,
\] 
while the angular momentum $J$ of the total system, 
that is, the spin of a highly excited meson, is 
$J=pL+S\approx pL$ with quark spin $S$. 
Therefore we have 
\[
H\approx{2Jc\over L}+\sigma L^n.
\] 
The stable energy state is realized by the 
conditon $\partial H/\partial L=0$, which gives 
the stable separation $L=(2Jc/n\sigma)^{1/(n+1)}$. 
Substituting this value back to the Hamiltonian, 
we obtain the Regge trajectory of the excited mesons as 
\[
M_J^2\propto J^{2n\over n+1}.
\] 
The observed spectrum of excited mesons and baryons lying on 
linearly rising Regge trajectories selects 
the universal value $n=1$, which in turn gives the linear potential 
$V(L)=\sigma L$ of quark confinement.

So, in classical as well as in quantum theories we only have 
the simplest exponents possible of $n=-1, 1,$ and $2$ operating 
for the dynamical force potentials in nature.

Now, the Dirac equation was found to explain 
the quantum mechanical motion of relativisitic electrons, 
in other words, to explain quantum and relativistic effects 
of fermionic degrees. 
The solutions of the equation not only explained remarkably 
the existence of electrons with negative charge, 
but also predicted the existence in nature of anti-electrons, 
nowadays called positrons, with degenerate mass 
and positive charge against their counterparts, electrons. 
No positrons had been found yet then, 
but only protons, neutrons, electrons and photons were around for study. 
Even Dirac himself in the beginning hesitated to accept 
the negative energy solutions of the equation 
predicting the existence of anti-electrons in nature, 
but rather tried to interpret them as 
describing the proton states of positive charge with some breaking 
of mass degeneracy.
Carl D. Anderson's experimental finding of a positron 
revealed here again 
that nature favors the beauty of Dirac equations with solid 
theoretical foundations based on 
the matter-antimatter symmetry, 
rather than the distorted interpretations of 
their solutions.

With this digestion of \lq known symmetries in nature\rq, 
let us go back to supersymmetry. 
Supersymmetry is a conjectured symmetry between fermions and bosons, 
and is inherently quantum mechanical since 
the very concept of fermions is quantum mechanical.
According to Witten~\cite{witt}, 
supersymmetry is an updating of 
special relativity to include fermionic as well as bosonic 
symmetries of spacetime. 
In developing relativity, Einstein assumed that the spacetime
coordinates were bosonic; fermions had not yet been discovered.
In supersymmetry the structure of spacetime is enriched by the 
presence of fermionic as well as bosonic coordinates:
\[
{\rm Superspace}\ X=\{x^\mu,\theta_\alpha\}.
\]
Supersymmetry, acting on the superspace coordinates, makes 
the following transformation:
\[
x^\mu\rightarrow x^\mu+i\bar\epsilon\gamma^\mu\theta
\]
\[
\theta_\alpha\rightarrow\theta_\alpha+\epsilon_\alpha.
\]
Experimental discovery of supersymmetry would be the 
beginning of probing the quantum structure of space and time, 
and my view is that the signal of supersymmetry 
might be the first appearance 
of \lq New Physics\rq.

Supersymmetry with supercharges are given as an extended 
algebra of Poincare group. 
It is based on solid theoretical foundations. 
\[ 
\{Q_a^A, \bar Q_{\dot b B}\}
=2\sigma_{a\dot b}^\mu P_\mu\delta^A_B,\] 
\[
\{Q_a^A, Q_b^B\}
=\{\bar Q_{\dot a A}, \bar Q_{\dot b B}\}=0,
\] 
\[
[P_\mu, Q_a^A]
=[P_\mu, \bar Q_{\dot a A}]=0,
\] 
\[
[Q_a^A,M_{\mu\nu}]={1\over 2}
(\sigma_{\mu\nu})_a^b Q_b^A,\quad
\bar [Q_{\dot a A},M_{\mu\nu}]=-{1\over 2}
(\sigma_{\mu\nu})_{\dot a}^{\dot b}\bar Q_{\dot b A},
\]
\[
\Bigl\{{\rm Poincare\ group\ with}\ P_\mu, M_{\mu\nu}\Bigr\}.
\]
It predicts the existence of \lq sparticles\rq\  
with degenerate masses against \lq particles\rq. 
On Earth we have not yet found those sparticles of low mass. 
Somewhere in the Cosmos there may be stars made up only of sparticles, 
or there may be sparticles of high mass in nature. 
In the last possibilitiy which we particle physicists are 
looking after, there must be some kind of supersymmetry-breaking. 
The breaking may be very cleverly and tactfully operating, 
maybe according to the yet-to-be-discoverd mechanism of 
supersymmetry-breaking,  
not just in the way of the spontaneous breaking of 
gauge symmetry of the Unified Theory. 
Who knows?

My belief is that Nature inherently loves Symmetry 
even if it necessarily exhibits some heavy-handed touch 
of Symmetry-Breaking.

So much for the physics part of my concluding remarks.

Now I must proceed to my final but most important role:
On behalf of the organizers of the Hiroshima Intenational Symposium 
I would like to express our sincere thanks to all the 
excellent speakers from abroad and from our country. 
Also, together with all participants I would like to extend 
our warm thanks first to the chairperson, Professor Jiro Kodaira, 
who has planned this conference so skillfully, 
and also to the other two local organizers, 
Professor T. Ohsugi and Dr. T. Oonogi. 
We attendants have enjoyed this Hiroshima meeting very very much.

Finally, but not the least, we wish to express our grattitude 
first to the Secretariat of all the graduate students of 
Hiroshima University including in particular Y. Kiyo, T. Nanno, 
H. Tochimura and K. Ishikawa, 
and also to the Secretaries, E. Masuda, M. K$\bar{\rm o}$no 
and M. Kubokawa of Physics Department, 
without whose help this Conference would not have been 
that fruitful, enjoyable, or that comfortable.

Thank you very much for your kind attention. 
See you soon again.

\vspace{2mm}

\begin{flushright}
October 29, 1997, in Hiroshima
\end{flushright}

\end{document}